\documentclass[12pt]{article}
\usepackage[margin=1in]{geometry}
\usepackage{setspace}

\usepackage{float}
\usepackage{amsfonts}

\usepackage{amssymb}
\usepackage{latexsym}
\usepackage{graphicx}
\usepackage[english]{babel}
\usepackage[font={small}]{caption}

\usepackage{amsfonts}
\usepackage{latexsym}
\usepackage{graphicx}
\usepackage[english]{babel}
\usepackage{tikz}

\begin{document}
\input epsf

\def\p{\partial}
\def\h{{1\over 2}}
\def\be{\begin{equation}}
\def\bea{\begin{eqnarray}}
\def\ee{\end{equation}}
\def\eea{\end{eqnarray}}
\def\d{\partial}
\def\la{\lambda}
\def\eps{\epsilon}
\def\bb{\bigskip}
\def\mm{\medskip}
\newcommand{\dm}{\begin{displaymath}}
\newcommand{\edm}{\end{displaymath}}
\renewcommand{\b}{\tilde{B}}
\newcommand{\gm}{\Gamma}
\newcommand{\ac}[2]{\ensuremath{\{ #1, #2 \}}}
\renewcommand{\ell}{l}
\newcommand{\z}{\ell}
\newcommand{\newsection}[1]{\section{#1} \setcounter{equation}{0}}
\def\bb{$\bullet$}
\def\Qbar{{\bar Q}_1}
\def\QPbar{{\bar Q}_p}

\def\q{\quad}

\def\bn{B_\circ}

\let\a=\alpha \let\b=\beta \let\g=\gamma \let\d=\delta \let\e=\epsilon
\let\c=\chi \let\th=\theta  \let\k=\kappa
\let\l=\lambda \let\m=\mu \let\n=\nu \let\x=\xi \let\r=\rho
\let\s=\sigma \let\t=\tau
\let\vp=\varphi \let\vep=\varepsilon
\let\w=\omega      \let\G=\Gamma \let\D=\Delta \let\Th=\Theta
                     \let\P=\Pi \let\S=\Sigma

\def\h{{1\over 2}}
\def\t{\tilde}
\def\r{\rightarrow}
\def\nn{\nonumber\\}
\let\bm=\bibitem
\def\Kt{{\tilde K}}
\def\b{\bigskip}

\let\p=\partial

\newcommand\blfootnote[1]{%
  \begingroup
  \renewcommand\thefootnote{}\footnote{#1}%
  \addtocounter{footnote}{-1}%
  \endgroup
}

\newcounter{daggerfootnote}
\newcommand*{\daggerfootnote}[1]{%
    \setcounter{daggerfootnote}{\value{footnote}}%
    \renewcommand*{\thefootnote}{\fnsymbol{footnote}}%
    \footnote[2]{#1}%
    \setcounter{footnote}{\value{daggerfootnote}}%
    \renewcommand*{\thefootnote}{\arabic{footnote}}%
    }

\begin{flushright}
\end{flushright}
\vspace{20mm}
\begin{center}
{\LARGE Are there echoes of gravitational waves?\daggerfootnote{Essay awarded an Honorable Mention in  the Gravity Research Foundation 2022 Awards for Essays on Gravitation.}
 }

\vspace{18mm}
{\bf Bin Guo$^1$ and Samir D. Mathur$^{2}$
\\}


\vspace{10mm}

$^{1}$Institut de Physique Th\'eorique,
	Universit\'e Paris-Saclay,
	CNRS, CEA, \\ 	Orme des Merisiers, Gif-sur-Yvette, 91191 CEDEX, France
	
E-mail: bin.guo@ipht.fr

\b

$^{2}$Department of Physics, The Ohio State University, Columbus,
OH 43210, USA

E-mail: mathur.16@osu.edu

\b

\b

\vspace{4mm}
\end{center}
\vspace{10mm}
\thispagestyle{empty}
\begin{abstract}

In several approaches to evading the information paradox, the semiclassical black hole is replaced by an Exotic Compact Object (ECO). It has been conjectured that gravitational waves emitted by the merger of ECOs can reflect off the ECOs and produce a detectable `echo'.  We argue that while a part of the wave can indeed reflect off the surface of an ECO, this reflected wave will get trapped by a {\it new} closed trapped surface produced by its own backreaction. Thus no detectable signal of the echo will emerge to infinity.  The only assumption in this analysis is that causality is maintained to leading order in gently curved spacetime. Thus if echoes are actually detected,  then we would face  a profound change in our understanding of physics.

\end{abstract}
\vskip 1.0 true in

\newpage
\setcounter{page}{1}

\doublespace


The observation of gravitational waves (GWs) has opened up an exciting new window in astronomy. What is perhaps even more exciting is that these waves could tell us something very nontrivial about the behavior of quantum gravity.

Consider two Schwarzcshild black holes of  mass $M$ each  that spiral in towards each other, emitting gravitational waves. In classical general relativity, the horizon is a vacuum region, with light cones that point `inwards'. Thus in the computation of GWs we place purely ingoing boundary conditions for these waves at the horizons.

But assuming such a vacuum horizon leads to a problem: in quantum theory, entangled pairs are created at the horizon,  and we run into Hawking's  information paradox \cite{hawking}. This problem is very robust: no small change to the dynamics at the horizon can resolve the paradox \cite{cern}. In string theory one finds that quantum gravitational effects  modify the entire interior of the classical hole, yielding a horizon sized object called a `fuzzball' \cite{fuzzballs}. The fuzzball radiates from its surface like a normal body; this resolves the paradox. Other approaches to the paradox have also sought to replace the classical hole by an ECO  --  an Exotic Compact object (for a review see \cite{Cardoso:2019rvt}). 

But with an ECO, we have the possibility that the part of the gravitational wave that falls towards one of the holes will reflect off the surface of the ECO, and produce an `echo' in the GW signal here on earth. Some have argued that we already see evidence of such echoes \cite{afshordi} while others have attributed the observed features to noise (for a review see \cite{Abedi:2020ujo}).  

Observations of GWs are likely to get dramatically better over the coming decade. One must therefore develop a good understanding of the  theoretical predictions regarding echoes.  In this article we will analyze the relevant physical effects, and find that the story has several interesting twists and turns. In brief, we find

(i) It is possible to make a model for an ECO where a significant fraction of the wave  reflects off the ECO.

(ii) The reflected wave, however, gets swallowed by a new closed trapped surface that forms outside the old horizon, due to the backreaction of the energy of the incoming wave.  Thus we should {\it not} be able to see this reflected wave in the GW signal received here on earth.

(iii) The only assumption in obtaining (ii) is that causality holds to leading order in gently curved regions of spacetime. Thus if we {\it do} see echoes, then we would learn something very nontrivial and unexpected about how quantum gravity behaves around black holes.   

Let us take each of these points in turn.

\b

{\bf (i) Can an ECO reflect a gravitational wave?}

\b

At first it might look obvious that an object with a surface like an ECO would be able to reflect a GW falling on it. After all,  the moon reflects the light falling on its surface. 

But an ECO is expected to be very compact; i.e., its surface should be a very small proper distance $s_{surface}$ outside the  radius $R=2GM$. If such were not the case, then we would simply be describing an normal body like a neutron star, and not an exotic compact object that is supposed  to replace a black hole. 

Consider a null ray that tries to head our from the surface of the ECO. If the ray is headed radially outwards, it escapes to infinity. But if it starts off at an angle $\theta$ to the radial direction, with
\be
\theta \gtrsim {s_{surface}\over R}
\ee
then it will arc back and fall onto the ECO instead of escaping to infinity. If we take $s_{surface}\sim l_p$ (the plank length) and $R\sim 3\, {\rm Km}$, then only null rays within a solid angle $\approx 10^{-76}$ around the radial direction will escape to infinity \cite{Guo:2017jmi}. 

This poses an issue. Suppose light falls onto an ECO which has a rough surface like the moon. Then the scattered ray would be more or less isotropically distributed over a solid angle $2\pi$, and a negligible fraction would esacpe back out to infinity. The same effect can be seen using waves. An incoming wave with $\lambda \sim R$ becomes a very short wavelength wave near the ECO due to redshift. Reflecting off a rough surface converts this wave to one with angular harmonics of order $l\sim R/s_{surface}$ and a correspondingly smaller energy in radial motion. These high $l$ waves cannot escape the angular momentum barrier of the Schwarzschild metric, so a negligible strength reaches infinity.

One could avoid the above conclusion by saying that the surface of the ECO is smooth like a mirror, so an incoming wave suffers `specular reflection' and heads back out to infinity.  Note however that a wave with $\lambda \sim R$ at infinity becomes a wave with $\lambda_{local}\sim l_p$ when it reaches to a proper distance $l_p$ from the surface $R=2GM$. Thus the reflecting surface would have to be smooth at the planck scale. Such smoothness is unnatural in view of the fact that the ECO should presumably have an entropy equal to the Bekenstein value $A/4G$, which is equivalent to one bit of data per planck sized plaquette on the horizon. A featureless surface would manifest no entropy. We could still argue that the interior of the ECO possesses entropy while the surface reflects like a mirror; however such a model does not look very natural.

The final possibility is that the surface of the ECO is not of a fixed shape with a given roughness, but rather is a fluid that can respond by undergoing collective oscillations when perturbed by  the incoming wave. In \cite{Chua:2021sew} it was shown that an electromagnetic wave can indeed reflect in a specular fashion off a plasma of electrons and positrons that is expected to surround an ECO radiating at the Hawking temperature. The question is: can a gravitational wave also suffer specular reflection from a suitable medium?

\begin{figure}
\centering
        \includegraphics[width=14cm]{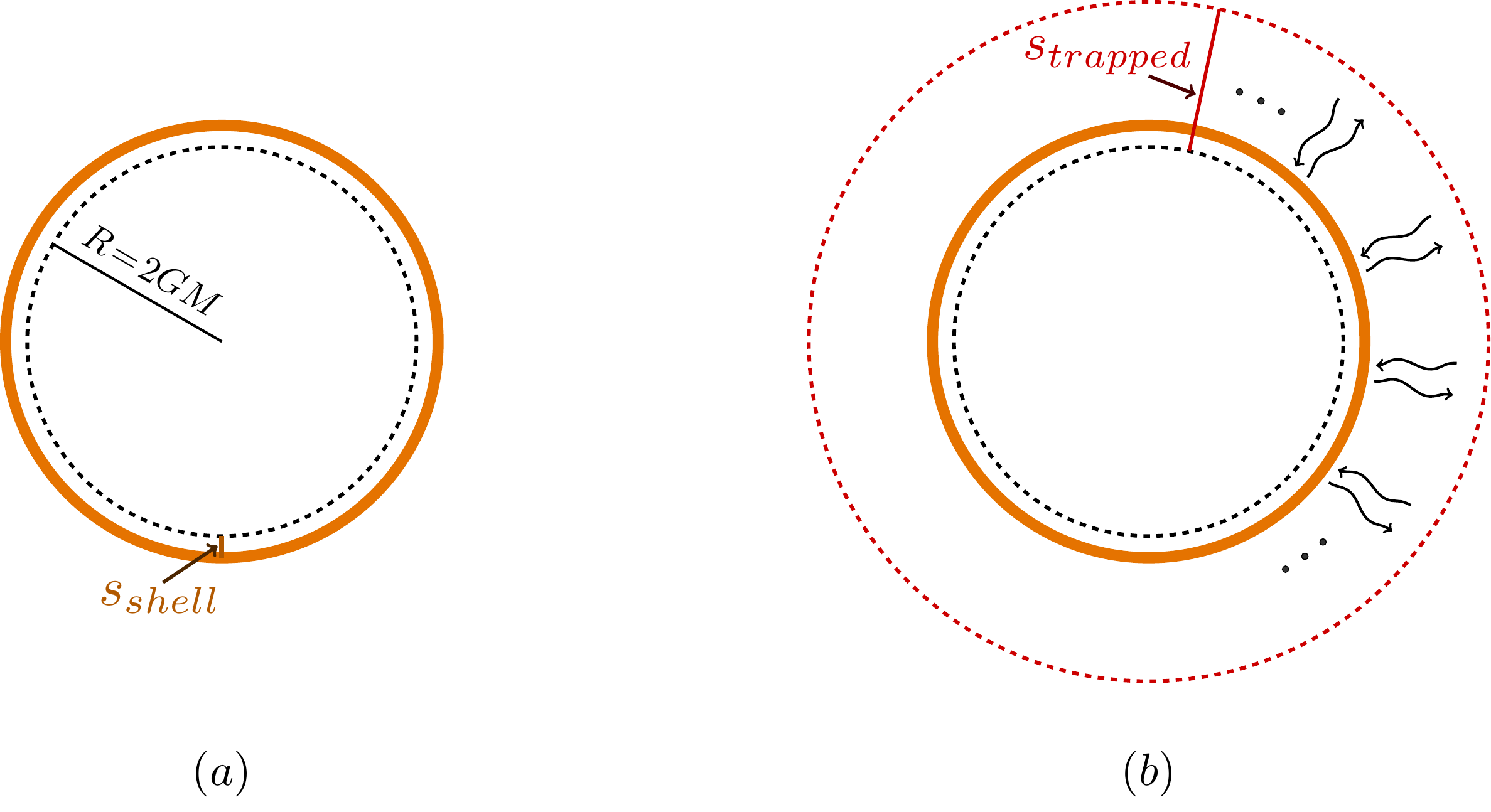}
\caption{(a) The exotic compact object (ECO) modelled as having a shell of perfect fluid a small distance $s_{shell}$ outside $r=2GM$. (b) An infalling gravitational wave can scatter back from this shell, but the scattered wave cannot escape the closed trapped surface that forms a distance $s_{trapped}$ outside $r=2GM$. }
\label{figone}
\end{figure}

Let the ECO have a thin shell of perfect fluid,  at a proper distance $s_{shell}$ from the radius $R=2GM$ (fig.\ref{figone}(a)). 
We take this fluid to have density $\rho$, pressure $p$, with $p$ of the same order as $\rho$. Since we are describing a model for the black hole rather than, say, a neutron star, we must have
\be
s_{shell}\ll R
\label{one}
\ee
The energy of this shell is $M'\approx 4\pi R^2 \rho$ in the local orthonormal frame at the location of the shell. It may seem that we must have $M'<M$, but there is a redshift $\sim R/s_{shell}$ at the location of the shell. Due to this, we have 
\be
M'\lesssim  \left ({R\over s_{shell}}\right) M
\label{two}
\ee
as the condition that the energy of the shell be less than the total mass of the ECO. 

We compute the reflection of an infalling GW off such a shell of perfect fluid. 
We find that a gravitational wave of amplitude $h_{incident}$ produces a reflected wave with amplitude
\be
h_{reflected}\sim  \left ( {GM' s_{shell}\over R^2}\right ) h_{incident}
\label{three}
\ee
The computation of the precise reflection amplitude will be presented elsewhere, but we outline the steps in the derivation that   yield (\ref{three}) as a rough estimate. First consider the length scales in the problem. Since most of the radiation is produced in the late stages of inspiralling, the distance between the holes $d$ and the wavelength $\lambda$ of the radiation are all $\sim R$, the radius of each hole.  We first rewrite the GW produced by the inspiralling in terms of spherical waves around one of the holes. The lowest tensor spherical harmonic is $l=2$, which has an amplitude $(R/\lambda)^3$ to penetrate the angular momentum barrier and reach the region close to shell. Since $\lambda\sim R$, we set this factor to order unity in our estimate.  When the wave reaches the shell, $h_{ab}$ distorts the shell of perfect fluid to produce a perturbed source with $\delta T_{ab} \propto h_{ab}$. This perturbed energy-momentum in turn produces an emitted wave, half of  which heads in the outwards direction. If this outward wave were not impeded by anything else, it would  penetrate the potential barrier and emerge to infinity; this barrier penetration factor is again order unity.  

In this computation we have assumed that the perfect fluid is able to come to equilibrium faster than the period of the wave that is perturbing it. The GW has $\lambda\gtrsim R$, so at a planck distance from the horizon its local wavelength is $\gtrsim l_p$. Thus the time for the fluid to come to equilibrium is required to be $\gtrsim l_p$; a response time like this is at least not ruled out by any principle of quantum gravity.

Using the upper limit in (\ref{two}) in (\ref{three}) gives 
\be
h_{reflected}\sim h_{incident}
\label{four}
\ee
The total luminosity of a wave depends not only on its amplitude $h_{ab}$ but also on the surface area of the region that it is emitted from. We have noted that the  size $d$ of the region emitting the original GW is $d \sim R$, while the size of the region emitting the reflected wave is also $\sim R$.  Thus (\ref{four}) suggests that the reflected wave will give an echo signal  that is not parametrically smaller than the signal of the primary GW.

\b

{\bf (ii) The effect of backreaction}

\b

It may appear from the above that we can in fact make a model of the ECO which will reflect the infalling GW and thereby create an echo in the signal observed on earth. But here we find another twist in the story. 

 Let a pulse of the  wave falling into the hole have energy $E$. The total energy of the ECO + pulse is now $M+E$. With this mass, a closed trapped surface (i.e. a new apparent horizon) will form at a proper  distance $s_{trapped}$ outside the radius $r=2GM$.   In the approximation where $s_{trapped}\lesssim R$. We have
\be
s_{trapped}\approx 4GM^\h E^\h
\label{five}
\ee
If 
\be
s_{shell} < s_{trapped}
\label{seven}
\ee
then the pulse that is reflected from our fluid shell  will not be able to emerge out to infinity (fig.\ref{figone}(b)). The pulse will be swallowed into the ECO, and its energy will then leak out slowly at the rate of Hawking radiation.  Thus there will be no observable echo signal  here on earth.

Now we note that the energy $E$ falling onto each hole is not parametrically smaller than $M$. When the two holes merge, a significant fraction of their energy goes into gravitational waves. Further since the separation $d$ between the orbiting holes is the same order as the radius $R$ of each hole,  a significant fraction of the wave does fall towards each hole. Suppose $E=.01 M$. Then for a solar mass hole with $R\sim 3 \, {\rm Km}$, we get $s_{trapped}\sim 600 \, m$. On the other hand for an ECO we have $s_{shell}\ll R$ (eq.(\ref{one})); for example with fuzzballs, it has been argued that the structure extends only order planck length $l_p$ outside $R=2GM$. Thus quite generally, we expect that with the above picture of the dynamics, we will have the situation (\ref{seven}) and the ECOs will not be able to send a reflected wave out to infinity.

One could ask the converse question: how weak does the gravitational wave have to be in order that the distance $s_{trapped}$ describing its backreaction is less than $s_{shell}$? In \cite{Guo:2017jmi,flaw} it was found that an incident wave with wavelength $\lambda \sim R$ containing order unity number of gravitons gives $s_{trapped}\sim l_p$. Thus if $s_{shell}\sim l_p$, the situation  (\ref{seven}) can be avoided only for GWs composed of just a few gravitons. Such weak waves will  not be detectable here on earth.

In short, while the infalling wave can indeed reflect off a suitably constructed ECO, the reflected wave will be trapped by the new apparent horizon that is created by its own energy, so the echo cannot escape to infinity.

\b

{\bf (iii) Does the closed trapped surface have to form?}

\b

At this point one may  ask the following question about the above analysis.  We have already assumed that quantum gravity effects in our theory act in a manner where they convert the semiclassical hole with vacuum horizon to an ECO with no horizon. Thus quantum gravity effects have produced an order unity change to our semiclassical picture at radii $r\sim R=2GM$. But in that case, how can we be sure  that the semiclassical argument for formation of a closed trapped surface at the location (\ref{five})  is still correct?   This is where we come to  the crucial constraint of causality.

Consider a black hole of mass $M$ that is created by $N$ particles, each moving radially inwards at the speed of light. No causal signal can travel from any of these particles to any other particle. Thus if causality holds in our theory, each particle must move exactly as it would if the other particles were not present. In particular, each particle will move uneventfully through the radius $r=2GM$, creating behind them a closed trapped surface at $r=2GM$. The formation of a  closed trapped surface was an essential part of  the above argument that the echo will not escape to infinity.

But if all this part of black hole dynamics is forced by causality, then where do we alter the semiclassical structure of the hole to evade the information paradox? In \cite{vecro} it was explained how a collapsing shell transforms to a fuzzball without violating causality anywhere. Let us recall this transformation process.

 String theory possesses {\it extended} structures -- fuzzballs -- with radius $R\sim 2GM$, for each value of $M$.  The number of these fuzzballs is large -- $Exp[S_{bek}(M)]$. We then find that the gravitational vacuum contains a significant component consisting of {\it virtual} fluctuations of fuzzball-type configurations: the natural suppression of the configurations with large $M$ is offset by the large degeneracy of these configurations. 
   
 From our constructions of fuzzballs, we know several properties of this  component of the gravitational vacuum. These virtual fluctuations are {\it extended}; a virtual fluctuation with energy $M$ has a size $\sim R_v\sim GM$. They are also {\it compression resistant}: they are local minima of the energy, so it costs extra energy to compress or stretch them. We call these  fluctuations {\it vecros}:  virtual extended compression-resistant objects.
 Because of this vecro component, the vacuum of quantum gravity is like a system at a second order phase transition: we have correlations at all length scales caused by virtual fluctuations of fuzzball microstates. 

What is the effect of having the vecro component in the gravitational vacuum? Consider a region where the spacetime curvature radius is $\sim R_{curv}$, and suppose that this curvature persists over a region of size $\sim L$. If $L\ll R_{curv}$, there is no significant effect of vecros; each vecro fluctuation of radius $R_v$ deforms in size by a small fraction $\sim R_v/L$, and stabilizes (due to the compression resistance) at its new size. The deformation energy of these vecros are already encoded in the low energy effective action $(16\pi G)^{-1}\int d^4 x {\mathcal R}$. 

\begin{figure}[H]
\begin{center}
\includegraphics[scale=.55]{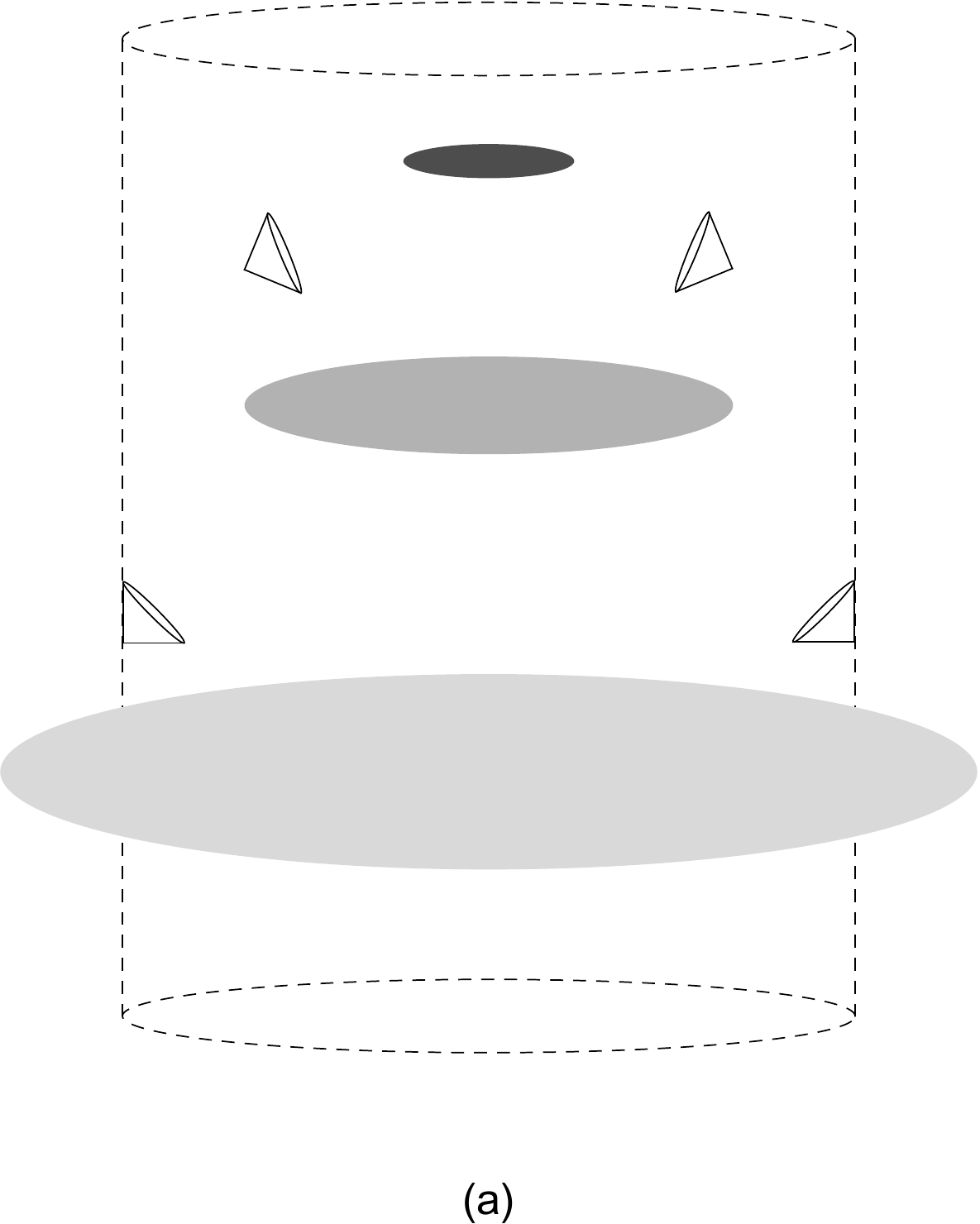}\hskip100pt
\includegraphics[scale=.55]{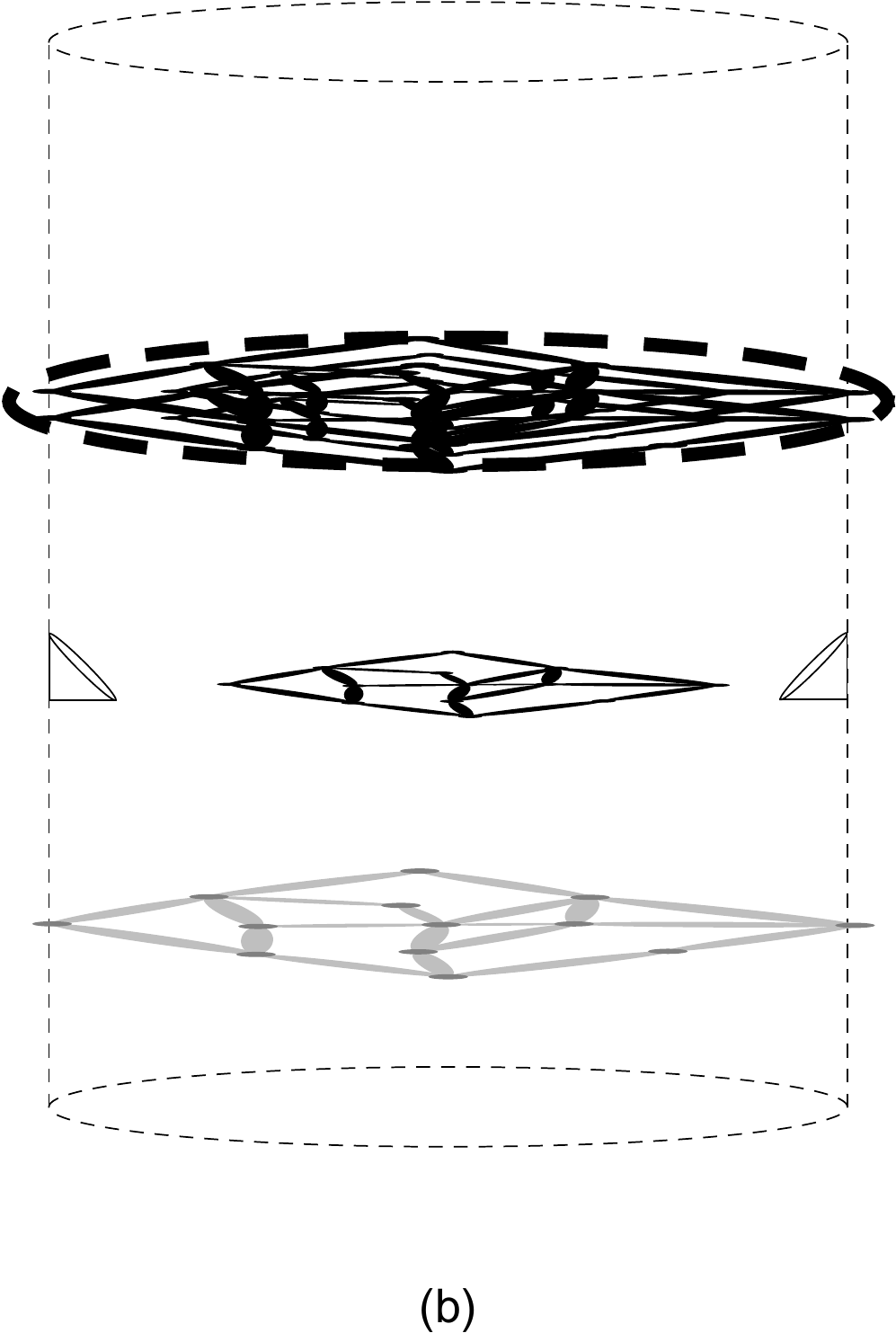}
\end{center}
\caption{(a) Semiclassical collapse: the gravitational collapse of a dust cloud generates a horizon; the light cones point inwards inside the horizon. (b) The modification due to vecros: at the bottom, a virtual fuzzball configuration (shown in grey); compression of this vecro leads to an on-shell excitation (shown higher up, in black); over a crossing timescale, such excitations generate a  horizon sized fuzzball (shown near the top).}
\label{fig20}
\end{figure}

But now suppose a closed trapped surface of radius $R$ forms around a point $r=0$; in this situation $L\sim R_{curv}$. The light cones in the region $r<R$ point purely inwards. So any vecro centered  around $r=0$ with radius $R_v<R$ suffers a monotonically increasing compression; it cannot stabilize at any size. When the compression factor becomes order unity, the corresponding deformation of the vecro part of the vacuum wavefunctional leads to the emergence of on-shell fuzzballs (fig.\ref{fig20}). 
 
The above process maintains causality of the underlying theory at all times. The extended nature of the virtual fuzzball fluctuations is crucial, as it allows the spacetime to detect the formation of a closed trapped surface and react by producing the fuzzball structure. {\it But note that a closed trapped surface must form first, by the causality argument above, before any nontrivial quantum gravity effects of fuzzballs can manifest themselves.} After that, in a few crossing times, the interior of the hole turns into a fuzzball. The fuzzball radiates from its surface like a normal body, thereby resolving the information paradox. 

\b

{\bf Summary} 

\b

Thus the essential issue boils down to the following. Suppose we assume that causality holds in the full quantum gravity theory, at least to leading order,  in regions where spacetime is gently curved ($R_{curv}\gg l_p$).  Then a closed trapped surface must first form around any infalling matter. Later, the vecro component of the vacuum can turn the interior of this trapped surface into a fuzzball without any violation of causality at any stage. The fuzzball does radiate energy from its surface that is $\sim l_p$ outside $r=2GM$, but this radiation is at the slow rate of Hawking evaporation. In the context of our analysis of gravitational waves,  this argument says  that in a theory with causality, the backreaction of the wave falling onto an ECO will cause the infalling wave to get trapped inside the closed trapped surface that it created, and an echo cannot be radiated out to infinity.

String theory does not violate causality, from all that we know at present. Many other approaches to quantum gravity also seek to maintain causality in the exact theory. Thus if we {\it do} see echoes, we would need to alter a very basic assumption about how quantum gravity works.  Observations may soon be able to find or rule out echoes in gravitational waves. Thus we are at a very exciting juncture in the science of gravitational waves.

\newpage

 \section*{Acknowledgements}

We would like to thank Oleg Lunin for many insights that went into this work. We would also like to thank Chris Hirata, Todd Thompson and David Weinberg for helpful comments. The work of B.G. is supported by the ERC grant 787320-QBH Structure. The work of S.D.M. is supported in part by DOE grant DE-SC0011726.

\end{document}